\documentclass[twocolumn,showpacs,superscriptaddress,10pt]{revtex4-2}
\usepackage{amsmath,amssymb,graphicx, subfigure, hyperref, braket, float}
\usepackage{blindtext}

\counterwithout{figure}{section}
\usepackage[export]{adjustbox}
\graphicspath{{figure/}}
\usepackage{graphicx, xcolor}
\usepackage[margin=.8in,text={10in,10.5in},centering]{geometry}
\usepackage{graphicx}
\usepackage{hyperref}
\usepackage{lipsum}
\usepackage{caption}
\usepackage{tikz}
\usepackage{needspace} 
\usepackage{bbold}
\usepackage{pgf}
\usepackage{ragged2e} 
\usepackage{booktabs}

\usepackage{bbold} 
\allowdisplaybreaks
\usepackage{array}

\allowdisplaybreaks
\hypersetup{ colorlinks=true,linkcolor=red,citecolor=blue,urlcolor=magenta}
\begin{document}
	
	\title{Multimode phonon-mediated enhancement of entanglement and competing synchronization in cavity magnomechanics}
	
	\author{Z. Imara} 
	\address{Laboratory of R\&D in Engineering Sciences, Faculty of Sciences and Techniques Al-Hoceima, Abdelmalek Essaadi University, Tetouan, Morocco}
	\author{Jia-Xin Peng} \email{JiaXinPeng@ntu.edu.cn} \thanks{(Corresponding Author)}
	\address{School of Physical Science and Technology, Nantong University, Nantong 226019, People's Republic of China}
	\author{E. K. Berinyuy} 
	\address{Department of Physics, Faculty of Science, University of Yaounde I, P.O.Box 812, Yaounde, Cameroon}
	\author{S.K. Singh} \email{singhshailendra3@gmail.com}\thanks{(Corresponding Author)}
	\address{Department of Physics, Akal University, Talwandi Sabo, Bathinda, Punjab 151302, India}
	\author{A. El Allati}
	\address{Laboratory of R\&D in Engineering Sciences, Faculty of Sciences and Techniques Al-Hoceima, Abdelmalek Essaadi University, Tetouan, Morocco}
	
\begin{abstract}

The generation of quantum correlations in hybrid quantum systems remains a central challenge due to the intrinsic limitations of linear interactions. In cavity magnomechanical platforms, the cavity-magnon coupling gives rise to hybridized cavity-magnon polaritons (CMPs). However, as a beam-splitter-type interaction, it does not by itself generate entanglement between the polariton modes in the absence of additional nonlinear or parametric processes. Here, we propose a mechanism based on multimode phonon mediation, in which multiple vibrational modes act as parallel scattering channels that couple the polaritons through Stokes and anti-Stokes processes. We show that , in the parameter regime explored here, the presence of multiple phonon modes leads to a monotonic enhancement of steady-state entanglement, thereby going beyond the limitations of conventional single-mode schemes. Furthermore, we demonstrate that quantum synchronization between the polariton modes originates from the same underlying scattering processes responsible for entanglement generation, yet exhibits an opposite scaling behavior with increasing phonon number for the phase quadrature, while the amplitude synchronization reveals collective squeezing that grows with the number of phonon channels. Our results provide new insights into the role of multimode interactions in shaping quantum correlations and establish a viable pathway for controlling entanglement and collective dynamics in hybrid magnomechanical platforms.

\end{abstract}
	
\maketitle
	
\section{Introduction}
	
	Cavity magnonics, which investigates the coherent interaction between microwave cavity photons and collective spin excitations (magnons) in magnetic materials such as yttrium iron garnet (YIG), has emerged as a rapidly developing field in hybrid quantum systems \cite{Rameshti2022}. Early theoretical work predicted the possibility of achieving strong coupling between a nanomagnet and a microwave cavity field \cite{Soykal2010}, laying the foundation for this field. These predictions were subsequently confirmed experimentally through the realization of strong cavity-magnon coupling in high-quality YIG-based systems \cite{A4,A5,A6}. The exceptional properties of YIG, including its high spin density and low intrinsic dissipation, make it an ideal platform for exploring coherent light--matter interactions.
	
	In the strong coupling regime, the interaction between cavity photons and magnons leads to the hybridization of their modes, forming two new normal modes known as cavity-magnon polaritons (CMPs). These hybrid modes, commonly referred to as the upper and lower polariton branches, exhibit a characteristic mode splitting that directly reflects the coupling strength. Polaritons, as hybrid quasiparticles combining both light and matter degrees of freedom, provide a versatile platform for integrating distinct physical systems and exploring macroscopic quantum phenomena \cite{Hopfield1958,Deng2010}. They have enabled the observation of remarkable effects such as Bose-Einstein condensation, quantized vortices, and superfluidity, while facilitating the development of novel quantum devices, including spin memories, light-emitting diodes, and optical transistors \cite{Kasprzak2006,Ballarini2013}.

	Despite these advantages, a fundamental limitation arises from the nature of the cavity--magnon interaction. The coupling is typically governed by a linear excitation-exchange (beam-splitter-type) interaction, which does not inherently generate quantum entanglement between the hybridized CMP modes \cite{A2}. Since entanglement generally requires nonlinear parametric processes, external quantum resources, or engineered interactions, directly entangling the two CMPs remains a challenging task. Although numerous theoretical proposals have explored entanglement and coherence generation in cavity magnomechanical and related hybrid systems \cite{jia,B1,Yuan2020,Shang2024,Wu2021,PP1,PP2,PP3}, a comprehensive and experimentally feasible framework for generating robust entanglement between polariton modes via multimode interactions is still lacking.
	
	On the other hand, synchronization, defined as the emergence of a common frequency or a fixed phase relationship between interacting oscillators, has been extensively studied across various physical platforms. In quantum systems, synchronization has been investigated in cavity optomechanics \cite{C2,Garg2023,Ghosh2024}, molecular optomechanical systems \cite{Peng2025}, and, more recently, in cavity magnonic systems \cite{Ghildiyal2026,J1}. These studies have revealed rich nonlinear dynamics and provided new avenues for controlling collective behavior in hybrid quantum systems.
	Furthermore, extending hybridization to include additional degrees of freedom, such as phonons, leads to polaromechanical systems where polaritons are coupled to mechanical modes. Recent experiments have demonstrated strong coupling between polaritons and phonons in both exciton-photon-phonon and magnon-photon-phonon platforms \cite{Kuznetsov2023}. Such polaromechanical interactions enable coherent control of mechanical states via polaritonic modes and open new possibilities for engineering quantum correlations and information transfer across different subsystems. Despite these advances, the interplay between polariton entanglement and quantum synchronization remains largely unexplored. In particular, the role of collective mediation mechanisms and multi-mode interactions in generating and controlling entanglement between CMPs, as well as their connection to synchronization phenomena, has not yet been systematically investigated.
	
	In this work, we address this gap by proposing a theoretical scheme to generate entanglement between two cavity-magnon polariton modes mediated by multiple vibrational (phonon) modes in a magnomechanical system. Our approach builds upon the well-established cavity-magnon polariton platform~\cite{A4,A5,A6,Rameshti2022} and extends it to the multimode regime. Related multimode phonon-mediated schemes have been proposed in cavity optomechanics~\cite{R11,R12}. Here, the mechanical eigenmodes of a YIG sphere naturally have distinct frequencies, and the hybridization angle between the photon and magnon components provides an additional continuously tunable control parameter specific to the magnomechanical platform. In contrast to single-phonon entanglement protocols~\cite{A3}, where entanglement relies on a single scattering channel, the proposed multimode configuration is not merely a straightforward extension. Instead, it establishes a regime in which phonon modes act as independent and parallel scattering channels that collectively mediate correlations between the polaritons.
	In the parameter regime explored here, we observe a monotonic enhancement of the stationary bipartite polariton entanglement with the number of phonon channels. We further demonstrate that quantum synchronization between the polariton modes arises from the same underlying scattering processes yet exhibits an opposite scaling behavior with respect to the number of phonon channels in the phase quadrature, while the amplitude quadrature reveals collective squeezing.
	This reveals a nontrivial interplay between entanglement generation and quantum synchronization in multimode hybrid systems. Our results provide new insights into the interplay between hybridization, nonlinearity, and collective dynamics and pave the way for exploiting polariton-based platforms in quantum information processing and hybrid quantum technologies.

	\section{Model and Hamiltonian}
	\label{sec:model}
	
	\begin{figure}[t]	
		\begin{flushleft}
			\subfigure{\label{A4}\includegraphics[scale=0.53]{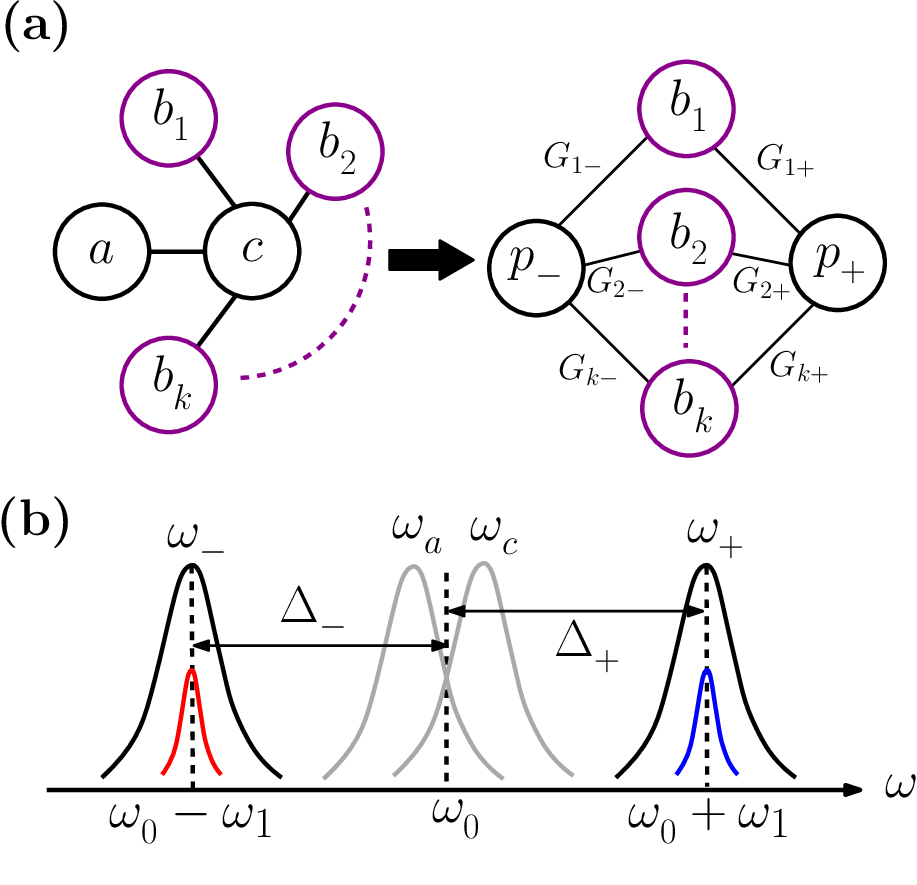}}
		\end{flushleft}
		\vspace{-1em}
		\captionsetup{justification=RaggedRight, singlelinecheck=false}
		\caption{{\small (a) Schematic of the multimode polaromechanical system. Strong coupling between the microwave cavity mode $a$ and magnon mode $c$ creates two hybrid cavity--magnon polariton (CMP) modes $p_+$ and $p_-$. Each polariton dispersively couples to $k$ mechanical modes ($b_j$, $j=1,\ldots,k$) via magnetostriction. (b) Frequency diagram. The magnon mode is driven at frequency $\omega_0$, and phonons scatter the driving photons onto Stokes ($\omega_0 - \omega_1$) and anti-Stokes ($\omega_0 + \omega_1$) sidebands. Setting the polariton frequencies resonant with the sidebands, $\omega_- \simeq \omega_0 - \omega_1$ and $\omega_+ \simeq \omega_0 + \omega_1$, enables entanglement between the two polaritons via collective phonon mediation.}}
		\label{fig:1}
	\end{figure}
	
	We consider a magnon-polaromechanical (magnon-PM) system \cite{A1, A2, A3}, as depicted in Fig.~\ref{fig:1}(a), which consists of a microwave cavity mode ($a$) strongly coupled with a magnon mode ($c$) leading to two magnon polaritons, which further couple to $k$ mechanical vibrational modes ($b_j$, $j = 1,\ldots,k$) via magnetostriction. Specifically, magnons are the collective excitations of a large number of spins in a yttrium-iron-garnet (YIG) sphere. They couple to microwave cavity photons via the magnetic dipole interaction \cite{A6,OO1} and to deformation phonons via the magnetostrictive force \cite{A7, A8, A9, A10}. Since the YIG sphere is typically of a large size yet much smaller than the microwave wavelength, the effect of radiation pressure is negligible \cite{B1,BA1}, and the frequencies of the vibrational modes are much lower than those of the cavity and magnon modes, yielding a dispersive magnomechanical interaction \cite{A2}. The total Hamiltonian of the system reads ($\hbar = 1$)
	\vspace*{-0.5em}
	\begin{eqnarray}
		\label{Eq.1}
		\mathcal{H} &=& \omega_a a^\dagger a + \omega_c c^\dagger c + g(a^\dagger c + a c^\dagger) + \sum_{j=1}^{k} \omega_j b_j^\dagger b_j 	\nonumber \\
		&+&\sum_{j=1}^{k}  G_{0j}(b_j + b_j^\dagger) c^\dagger c  + i\Omega_0(c^\dagger e^{-i\omega_0 t} - c e^{i\omega_0 t}),
	\end{eqnarray}
	where $a$, $c$, and $b_j$ are the annihilation operators of the cavity, magnon, and mechanical modes, respectively, satisfying $[O,O^\dagger] = 1$, and $\omega_a$, $\omega_c$, and $\omega_j$ are their resonance frequencies. The magnon frequency is tunable over a wide range by varying the external bias magnetic field $H_0$ via $\omega_c = \gamma_0 H_0$, where $\gamma_0/2\pi = 28$~GHz/T is the gyromagnetic ratio. The magnon-cavity coupling strength $g$ can be much larger than the cavity and magnon decay rates, $g > \kappa_a, \kappa_c$, entering the strong coupling regime and resulting in two hybridized magnon polaritons~\cite{A4, A5, A6}. The bare magnomechanical coupling rates $G_{0j}$ are typically small (e.g., $G_{0j}/2\pi \simeq 0.2$~Hz for a 250-$\mu$m-diameter YIG sphere~\cite{B1}). The Rabi frequency $\Omega_0 = \sqrt{5}\gamma_0\sqrt{N}B_0/4$~\cite{B1} characterizes the coupling between the magnon mode and its drive field with frequency $\omega_0$ and amplitude $B_0$, where $N = \rho V$ is the total number of spins, with $\rho = 4.22\times 10^{27}$~m$^{-3}$ being the spin density of the YIG and $V$ the volume of the sphere. Note that $\Omega_0$ is derived under the assumption of low-lying excitations, $\langle c^\dagger c \rangle \ll 2Ns$, where $s = 5/2$ is the spin number of the ground-state Fe$^{3+}$ ion in YIG~\cite{B1}.

	In the strong coupling regime forming two polaritons, the system becomes a polaromechanical system. It is thus convenient to express the Hamiltonian~\eqref{Eq.1} with the polariton operators $p_\pm$. 
	In the interaction picture with respect to $\hbar\omega_0(p_+^\dagger p_+ + p_-^\dagger p_-)$, the Hamiltonian reads
\begin{eqnarray}
	\label{Eq.2}
	\mathcal{H} &=& \Delta_+ p_+^\dagger p_+ + \Delta_- p_-^\dagger p_- + \sum_{j=1}^{k} \omega_j b_j^\dagger b_j
	\nonumber \\
	&+& \sum_{j=1}^{k} G_{0j} \textcolor{blue}{c^\dagger c} (b_j + b_j^\dagger) + i\Omega_0(\textcolor{blue}{c^\dagger - c}),
\end{eqnarray}
where $p_+$ and $p_-$ are the annihilation operators of the upper and lower polaritons, which are the hybridization of the cavity and magnon modes via the transformation
\begin{equation}
	\label{Eq.3}
	\begin{pmatrix} p_+ \\ p_- \end{pmatrix} = 
	\begin{pmatrix} \cos\theta & \sin\theta \\
		-\sin\theta & \cos\theta \end{pmatrix}
	\begin{pmatrix} a \\ c \end{pmatrix},
\end{equation}
satisfying the bosonic commutation relations $[p_\pm, p_\pm^\dagger] = 1$ and $[p_+, p_-] = 0$. Here,
\begin{eqnarray}
	\label{Eq.4}
	\theta = \frac{1}{2}\arctan\!\left(\frac{2g}{\Delta_{ac}}\right),
\end{eqnarray}
with $\Delta_{ac} = \omega_a - \omega_c$, characterizes the mixing ratio of photons and magnons in the polaritons, $\Delta_\pm = \omega_\pm - \omega_0$ are the polariton-drive detunings, and the eigenfrequencies are
\begin{eqnarray}
	\label{Eq.5}
	\omega_\pm = \frac{\omega_a + \omega_c \pm \sqrt{\Delta_{ac}^2 + 4g^2}}{2}.
\end{eqnarray}

	Including dissipations and input noises for each mode, we obtain the set of quantum Langevin equations (QLEs) for the system dynamics as 
	\begin{eqnarray}
		\dot{p}_+ &=& -i\Delta_+ p_+ - \kappa_+ p_+ - \delta\kappa\, p_- + \Omega_0\sin\theta 
		\nonumber \\
		&\quad& - i\sum_{j=1}^{k} G_{0j}(b_j + b_j^\dagger)\, c\,\sin\theta + \sqrt{2\kappa_+}\,p_+^{\rm in},
		\nonumber \\
		\dot{p}_- &=& -i\Delta_- p_- - \kappa_- p_- - \delta\kappa\, p_+ + \Omega_0\cos\theta 
		\nonumber \\
		&\quad& - i\sum_{j=1}^{k} G_{0j}(b_j + b_j^\dagger)\, c\,\cos\theta + \sqrt{2\kappa_-}\,p_-^{\rm in},
		\nonumber \\
		\dot{b}_j &=& -i\omega_j b_j - \kappa_j b_j - iG_{0j}\,c^\dagger c + \sqrt{2\kappa_j}\,b_j^{\rm in},
		\label{Eq.6}
	\end{eqnarray}
	where $\kappa_+ = \kappa_a\cos^2\!\theta + \kappa_c\sin^2\!\theta$ and $\kappa_- = \kappa_a\sin^2\!\theta + \kappa_c\cos^2\!\theta$ are the two polariton dissipation rates, and $\delta\kappa \equiv (\kappa_c - \kappa_a)\sin\!\theta\cos\!\theta$ signifies the dissipative coupling between the two polaritons arising from the unbalanced cavity and magnon dissipation rates. Here, $p_+^{\rm in} \equiv (\sqrt{2\kappa_a}\cos\theta\, a^{\rm in} + \sqrt{2\kappa_c}\sin\theta\, c^{\rm in})/\sqrt{2\kappa_+}$ and $p_-^{\rm in} \equiv (-\sqrt{2\kappa_a}\sin\theta\, a^{\rm in} + \sqrt{2\kappa_c}\cos\theta\, c^{\rm in})/\sqrt{2\kappa_-}$ are the input noise operators of the polaritons, $\kappa_j$ and $b_j^{\rm in}$ are the damping rate and input noise of the mechanical mode $b_j$, and all input noises are zero-mean and characterized by the correlation functions \cite{B2}: 
	\begin{eqnarray}
		\langle o^{\rm in}(t) o^{{\rm in}\dagger}(t') \rangle &=& [\bar{n}_o(\omega_o)+1]\,\delta(t-t'), \nonumber \\
		\langle o^{{\rm in}\dagger}(t)\, o^{\rm in}(t') \rangle &=& \bar{n}_o(\omega_o)\,\delta(t-t'),
	\end{eqnarray} 
	where $\bar{n}_o(\omega_o) = [\exp(\hbar\omega_o/k_BT)-1]^{-1}$ are the mean thermal excitation numbers, with $k_B$ the Boltzmann constant and $T$ the bath temperature.
	
	Using a strong coherent drive on the magnon mode, and due to its excitation-exchange interaction with the cavity mode, the polariton amplitudes become large, $|\langle p_+\rangle|, |\langle p_-\rangle| \gg 1$ at steady-state. Thus, one should linearize the QLEs in Eq.~\eqref{Eq.6} around the steady-state value by writing each mode operator as $o = \langle o\rangle + \tilde{o}$ ($o = p_+, p_-, b_1, \ldots, b_k$), and neglect the second-order fluctuation terms. The linearized QLEs describe the quantum fluctuations of the system using the quadrature fluctuation operators ($\tilde{X}_v,\tilde{Y}_v$), where $\tilde{X}_v=1/\sqrt{2}(\tilde{v}^\dagger+ \tilde{v})$ and $\tilde{Y}_v=i/\sqrt{2}(\tilde{v}^\dagger -\tilde{v})$ with ($v = p_+, p_-, b_1, \ldots, b_k$), and can be written in matrix form.
	\begin{equation}
		\dot{\mathcal{L}}(t) = \Lambda \mathcal{L}(t) + \mathcal{N}(t),
		\label{eq:10}
	\end{equation}
	where 
	$\mathcal{L} = (\tilde{X}_{p_+}, \tilde{Y}_{p_+}, \tilde{X}_{p_-}, \tilde{Y}_{p_-}  ,\tilde{X}_{b_1} ,\tilde{Y}_{b_1},\ldots, \tilde{X}_{b_k} ,\tilde{Y}_{b_k})^{\rm T}$ is the vector of quadrature fluctuations, $ \mathcal{N}(t) = (\sqrt{2\kappa_+} \tilde{X}_{p_+}^{\rm in}, \sqrt{2\kappa_+}\tilde{Y}_{p_+}^{\rm in}, \sqrt{2\kappa_-}\tilde{X}_{p_-}^{\rm in}, \sqrt{2\kappa_-}\tilde{Y}_{p_-}^{\rm in}, \sqrt{2\kappa_1}\tilde{X}_{b_1}^{\rm in},\\ \sqrt{2\kappa_1}\tilde{Y}_{b_1}^{\rm in}, \ldots, \sqrt{2\kappa_k}\tilde{X}_{b_k}^{\rm in}, \sqrt{2\kappa_k}\tilde{Y}_{b_k}^{\rm in})^{\rm T}$ is the vector of input noises expressed in quadrature form.
	Moreover, $\Lambda$ is the drift matrix, given by
	\begin{equation}
		\Lambda = \left(\begin{array}{cc|ccc}
			\mathcal{F}_+ & -\delta\kappa\mathbb{1}_{2\times2} & \mathcal{A}_{1+} & \cdots & \mathcal{A}_{k+} \\
		-\delta\kappa\mathbb{1}_{2\times2} & \mathcal{F}_- & \mathcal{A}_{1-} & \cdots & \mathcal{A}_{k-} \\
			\hline
			\mathcal{B}_{1+} & \mathcal{B}_{1-} & \mathcal{F}_1 & & \mathbf{0}_k \\
			\vdots & \vdots & & \ddots & \\
			\mathcal{B}_{k+} & \mathcal{B}_{k-} & \mathbf{0}_k & & \mathcal{F}_k
		\end{array}\right),
		\label{Eq:11}
	\end{equation}
	where $\mathcal{F}_\pm$, $\mathcal{F}_j$, $\mathcal{A}_{j\pm}$, and $\mathcal{B}_{j\pm}$ are $2\times 2$ matrices:
	\begin{align}
		\mathcal{F}_\pm &= \begin{pmatrix} -\kappa_\pm & \bar{\Delta}_\pm \\ -\bar{\Delta}_\pm & -\kappa_\pm \end{pmatrix}, \quad
		\mathcal{F}_j = \begin{pmatrix} -\kappa_j & \omega_j \\ -\omega_j & -\kappa_j \end{pmatrix}, 
		\label{eq:F_blocks} \\[4pt]
		\mathcal{A}_{j+} &= \begin{pmatrix} -\mathcal{G}_j\sin\theta & 0 \\ 0 & 0 \end{pmatrix}, \quad
		\mathcal{A}_{j-} = \begin{pmatrix} -\mathcal{G}_j\cos\theta & 0 \\ 0 & 0 \end{pmatrix},
		\label{eq:A_blocks} \\[4pt]
		\mathcal{B}_{j+} &= \begin{pmatrix} 0 & 0 \\ 0 & \mathcal{G}_j\sin\theta \end{pmatrix}, \quad
		\mathcal{B}_{j-} = \begin{pmatrix} 0 & 0 \\ 0 & \mathcal{G}_j\cos\theta \end{pmatrix}.
		\label{Eq:12}
	\end{align}
	Here, $\bar{\Delta}_+={\Delta}_+ + \sum_{j=1}^{k}2 G_{0j} \mathbf{Re}\langle b_j \rangle\sin^2\theta$ and $\bar{\Delta}_-={\Delta}_- + \sum_{j=1}^{k}2 G_{0j} \mathbf{Re}\langle b_j \rangle\cos^2\theta$ are effective detunings that include the frequency shift induced by the PM interaction, where $\langle b_j\rangle = -{G_{0j}|\langle c\rangle|^2}/({\omega_j-i\kappa_j})$, with $j = 1,\ldots,k$.
    The effective magnomechanical coupling strength associated with the mechanical mode $b_j$ is $\mathcal{G}_j = 2iG_{0j}\langle c\rangle$, where the steady-state magnon amplitude $\langle c\rangle$ is obtained from Eq.~\ref{Eq.3} as
    \begin{eqnarray}
		\langle c\rangle &=& \langle p_+\rangle \sin\theta + \langle p_-\rangle \cos\theta,
	\end{eqnarray}
	where 
	\begin{eqnarray}
		\langle p_+\rangle &=& -\frac{\delta \kappa \Omega_0 \cos\theta - i\Omega_0\sin\theta(\Delta_- - i\kappa_-)}{(\Delta_- - i\kappa_-)(\Delta_+ - i\kappa_+) + \delta\kappa^2}, \\
		\langle p_-\rangle &=& -\frac{\delta \kappa \Omega_0 \sin\theta - i\Omega_0\cos\theta(\Delta_+ - i\kappa_+)}{(\Delta_- - i\kappa_-)(\Delta_+ - i\kappa_+) + \delta\kappa^2},
	\end{eqnarray}
	
	Note that the drift matrix is given under the condition $|\Delta_\pm| \gg \kappa_\pm, \delta\kappa$, well satisfied in the experiments~\cite{A2, A1}, and is the optimal condition for quantum correlations in the system. Since in this regime, $\langle b_j\rangle \simeq -{G_{0j}|\langle c\rangle|^2}/{\omega_j}$, and the weak $G_{0j}$, the frequency shifts in the detunings are typically small, i.e., $|\bar{\Delta}_\pm - \Delta_\pm| \ll \Delta_\pm$, and can be safely neglected. Furthermore, $\langle c\rangle$ takes a simpler form
	\begin{equation}
		\langle c\rangle \simeq -i \left(\frac{\sin^2\theta}{\Delta_+} + \frac{\cos^2\theta}{\Delta_-}\right) \Omega_0,
		\label{eq:18}
	\end{equation}
	which is a pure imaginary number, leading to a real $\mathcal{G}_j$ and simplifying the structure of the coupling blocks. The $\sin\theta$ and $\cos\theta$ factors in $\mathcal{A}_{j\pm}$ and $\mathcal{B}_{j\pm}$ reflect the magnon weight in each polariton, which determines how the magnomechanical coupling $\mathcal{G}_j$ is distributed between the upper ($p_+$) and lower ($p_-$) PM couplings.
	
	Due to the linearized dynamics and the Gaussian nature of the quantum noises, the steady state of the quantum fluctuations is a continuous-variable $(2+k)$-mode Gaussian state, where $k$ denotes the number of phonon modes, which is completely characterized by a $(4+2k)\times(4+2k)$ covariance matrix (CM) $\mathcal{V}$ with entries $\mathcal{V}_{ij} = \frac{1}{2}\langle \mathcal{L}_i(t)\mathcal{L}_j(t) + \mathcal{L}_j(t)\mathcal{L}_i(t)\rangle$. The steady-state CM can be obtained by directly solving the Lyapunov equation \cite{B3}
	\vspace*{-0.2em}
	\begin{equation}
		\Lambda\,\mathcal{V} + \mathcal{V}\,\Lambda^{\rm T} = -\Gamma,
		\label{eq:19}
	\end{equation}
	where $\Gamma$ is the diffusion matrix, defined via $\Gamma_{ij}\,\delta(t-t') = \langle n_i(t)n_j(t') + n_j(t')n_i(t)\rangle/2$, and takes the form 
		\vspace*{-0.2em}
	\begin{equation}
		\Gamma = \Gamma^{\rm d} + \Gamma^{\rm o}.
	\end{equation}
	The diagonal part is
	\begin{eqnarray}
		\Gamma^{\rm d} = \Gamma_{p_+} \oplus \Gamma_{p_-} \oplus \Gamma_{b_1} \oplus \ldots \oplus \Gamma_{b_k},
		\label{eq:Eq17}
	\end{eqnarray}
	and the off-diagonal part $\Gamma^{\rm o}$ has nonzero entries only in the $4\times 4$ polariton block:
	\begin{equation}
		\Gamma^{\rm o}_{4\times4}  =  \tilde{\kappa}_p \left(\begin{array}{cc}
			\mathbf{0}_{2\times2} & \mathbb{1}_{2\times2}\\
			\mathbb{1}_{2\times2} & \mathbf{0}_{2\times2}  \\
		\end{array}\right),
	\end{equation}
	where $\Gamma_{p_\pm} = {\rm diag}[\tilde{\kappa}_\pm, \tilde{\kappa}_\pm]$, $\Gamma_{b_j} = {\rm diag}[\kappa_j(2\bar{n}_{b_j}+1),\kappa_j(2\bar{n}_{b_j}+1)]$, 
	with 
	\begin{eqnarray}
		\tilde{\kappa}_+&=& \kappa_a\cos^2\theta(2\bar{n}_a+1) + \kappa_c\sin^2\theta\,(2\bar{n}_c+1), \nonumber\\
		\tilde{\kappa}_-&=& \kappa_a\sin^2\theta(2\bar{n}_a+1) + \kappa_c\cos^2\theta(2\bar{n}_c+1),\nonumber\\
		\tilde{\kappa}_p&=& \sin\theta\cos\theta[-\kappa_a(2\bar{n}_a+1) + \kappa_c(2\bar{n}_c+1)].
	\end{eqnarray}
	Note that $\tilde{\kappa}_p$ vanishes for balanced dissipation rates $\kappa_a = \kappa_c$ at equal bath temperatures. This off-diagonal contribution arises from the cross-correlations between the polariton input noises due to the unbalanced decay channels, and is fully included in our numerical calculations. The analysis of steady-state properties requires that the real parts of all eigenvalues of $\Lambda$ be negative, which we verify numerically for each parameter set considered.

	To quantify the quantum correlations in the system, we adopt several measures. 
	The bipartite entanglement between two Gaussian modes is quantified by the logarithmic negativity \cite{B4,B5,B6,BB1,BB2}
	\begin{equation}
		E_N = \max\{0,\,-\ln(2\tilde{\nu}_-)\},
		\label{eq:EN}
	\end{equation}
	where $\tilde{\nu}_-$ is the smallest symplectic eigenvalue of the partially transposed covariance matrix (CM), obtained as 
	$\tilde{\nu}_- = \min |{\rm eig}[i\boldsymbol{\Omega}_2\tilde{\mathbf{V}}_4]|$. 
	Here $\boldsymbol{\Omega}_2 = \bigoplus_{j=1}^{2} i\sigma_y$ is the two-mode symplectic form, with $\sigma_y$ the Pauli-$y$ matrix. 
	The partially transposed CM is given by $\tilde{\mathbf{V}}_4 = \mathbf{P}_0\,\mathbf{V}_4\,\mathbf{P}_0$, where 
	$\mathbf{P}_0 = {\rm diag}(1,-1,1,1)$ implements the partial transposition at the level of covariance matrices. 
	The matrix $\mathbf{V}_4$ is the $4\times4$ reduced CM of the two subsystems and has the block structure
	\begin{equation}
		\mathbf{V}_4 =
		\begin{pmatrix}
			\mathbf{V}_{p_+} & \mathbf{V}_{p_+p_-} \\
			\mathbf{V}_{p_+p_-}^{\rm T} & \mathbf{V}_{p_-}
		\end{pmatrix},
		\label{eq:V4}
	\end{equation}
	where $\mathbf{V}_{p_+}$ and $\mathbf{V}_{p_-}$ are the $2\times2$ covariance matrices of the individual polariton modes and 
	$\mathbf{V}_{p_+p_-}$ contains their cross-correlations. 
	The matrix $\mathbf{V}_4$ is obtained from the full $(4+2k)\times(4+2k)$ CM $\mathbf{V}$ by tracing over the mechanical degrees of freedom. 
	A positive value $E_N>0$ certifies the presence of bipartite entanglement between the two modes.

    Beyond bipartite correlations, we also characterize \textit{genuine} tripartite entanglement among the two polaritons and each phonon mode $b_j$ through the minimum residual contangle~\cite{B8,B9}
	\begin{equation}
		\mathcal{R}_{\min}^{\tau} = \min_{(r,s,t)} 
		\left\{E_{\tau}^{r|st} - E_{\tau}^{r|s} - E_{\tau}^{r|t}\right\},
		\label{eq:Rtau}
	\end{equation}
	where $(r,s,t)$ denotes the permutations of the three-mode indices $(p_+,p_-,b_j)$, and $E_{\tau}^{u|v} \equiv [E_N^{u|v}]^2$ is the contangle~\cite{B11}.
	The $6\times6$ CM of the three-mode subsystem has the block structure
	\begin{equation}
		\mathbf{V}_6 =
		\begin{pmatrix}
			\mathbf{V}_{p_+} & \mathbf{V}_{p_+p_-} & \mathbf{V}_{p_+b_j} \\
			\mathbf{V}_{p_+p_-}^{\rm T} & \mathbf{V}_{p_-} & \mathbf{V}_{p_-b_j} \\
			\mathbf{V}_{p_+b_j}^{\rm T} & \mathbf{V}_{p_-b_j}^{\rm T} & \mathbf{V}_{b_j}
		\end{pmatrix},
		\label{eq:V6}
	\end{equation}
	where $\mathbf{V}_{p_+}$, $\mathbf{V}_{p_-}$, and $\mathbf{V}_{b_j}$ are the $2\times2$ covariance matrices of the individual modes, and the off-diagonal blocks contain their pairwise correlations. 
	For bipartitions where one subsystem contains two modes, the symplectic eigenvalue $\tilde{\nu}_-$ is obtained from 
	$\tilde{\nu}_- = \min |{\rm eig}[i\boldsymbol{\Omega}_3\tilde{\mathbf{V}}_6]|$, where
	\begin{equation}
		\boldsymbol{\Omega}_3 = \bigoplus_{j=1}^{3} i\sigma_y
	\end{equation}
	is the three-mode symplectic form. 
	The partially transposed CM is $\tilde{\mathbf{V}}_6 = \mathbf{P}\mathbf{V}_6\mathbf{P}$, where the partial transposition matrices for the three bipartitions are
	\begin{equation}
		\begin{aligned}
			\mathbf{P}_{p_+|p_-b_j} &= {\rm diag}(1,-1,1,1,1,1), \\
			\mathbf{P}_{p_-|p_+b_j} &= {\rm diag}(1,1,1,-1,1,1), \\
			\mathbf{P}_{b_j|p_+p_-} &= {\rm diag}(1,1,1,1,1,-1).
		\end{aligned}
		\label{eq:P_matrices}
	\end{equation}
	A nonzero $\mathcal{R}_{\min}^{\tau}>0$ certifies genuine tripartite entanglement among the two polaritons and the phonon mode~\cite{B8}.

Beyond entanglement, we quantify the quantum synchronization between the two polariton modes following the framework of Ref.~\cite{C2}. Defining the difference quadratures $\tilde{X}_- = (\tilde{X}_{p_+} - \tilde{X}_{p_-})/\sqrt{2}$, $\tilde{P}_- = (\tilde{P}_{p_+} - \tilde{P}_{p_-})/\sqrt{2}$, and more generally the arbitrary quadrature $X_\phi = \tilde{X}_-\cos\phi + \tilde{P}_-\sin\phi$, we introduce the synchronization measures 
\begin{equation} 
	S_c = \frac{1}{\langle \tilde{X}_-^2 \rangle + \langle \tilde{P}_-^2 \rangle}\,, \qquad S(\phi) = \frac{1}{2\langle X_\phi^2 \rangle}, 
	\label{Eq:28} 
	\end{equation}
where the variances are extracted from the covariance matrix $\mathcal{V}$. The measure $S(\phi)$ reduces to the amplitude synchronization $S_x = 1/(2\langle \tilde{X}_-^2 \rangle)$ for $\phi = 0$ and to the phase synchronization $S_p = 1/(2\langle \tilde{P}_-^2 \rangle)$ for $\phi = \pi/2$.
The Heisenberg principle restricts $0 \leq S_c \leq 1$, where $S_c = 0$ corresponds to fully unsynchronized modes and $S_c = 1$ to perfect synchronization. For the synchronization at an arbitrary quadrature, $0 < S(\phi) \leq 1$ indicates that synchronization is achieved, while $S(\phi) > 1$ reveals that the collective fluctuations between the two polaritons are squeezed below the vacuum level in the corresponding quadrature \cite{C2}.
    
	\section{Results and Discussion}
	\label{sec:results}

	\begin{figure*}[ht]
	\centering
	\includegraphics[scale=0.27]{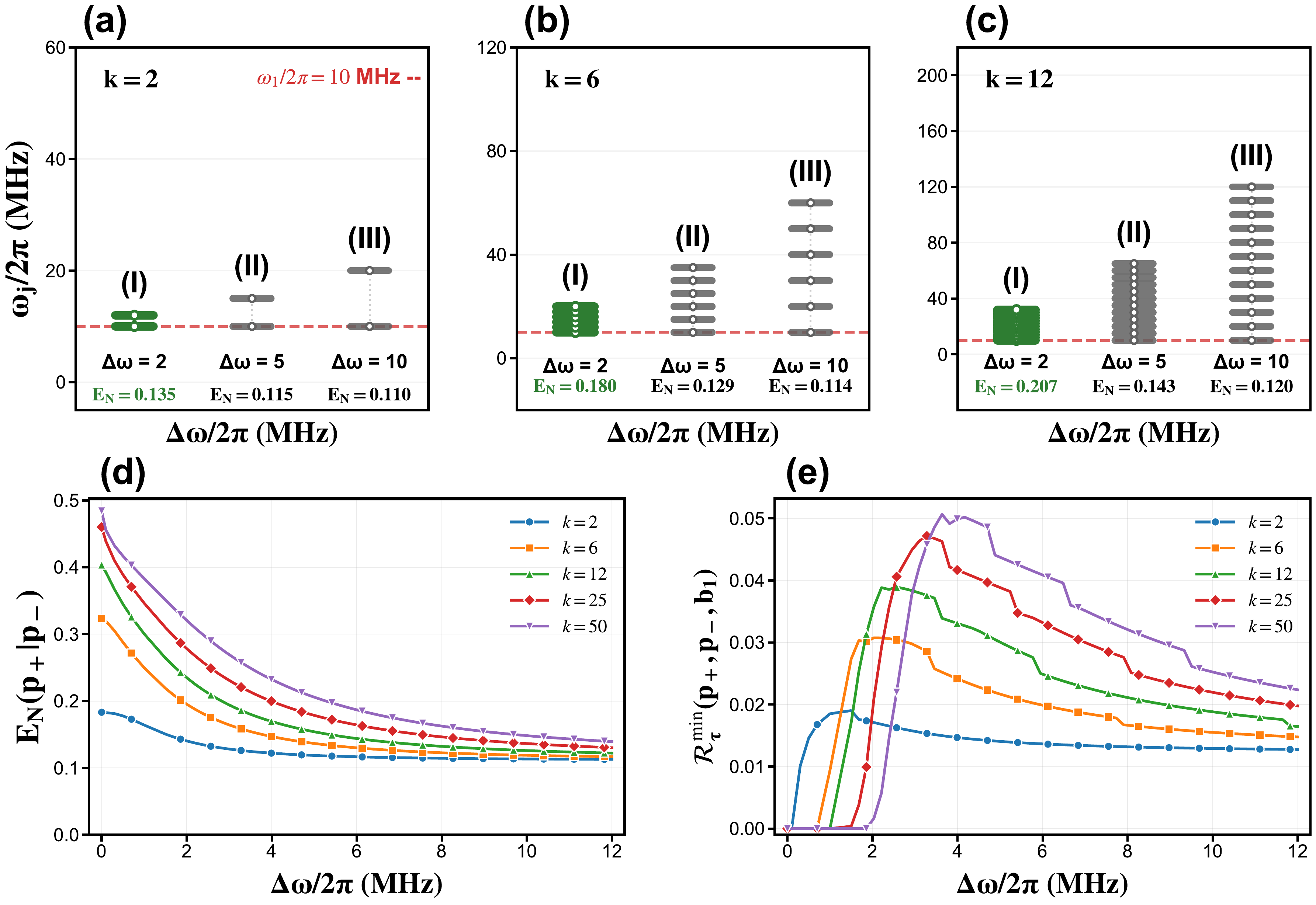}
	\captionsetup{justification=RaggedRight, singlelinecheck=false}
	\vspace{-0.2em} 
	\caption{{\small Effect of the phonon frequency spacing $\Delta\omega$ on the polariton-polariton entanglement and on the tripartite entanglement. (a)-(c) Spectral configurations for $k = 2$, $k = 6$, and $k = 12$ phonon modes, respectively. Three spacings are considered in each panel: $\Delta\omega/2\pi = 2$~MHz (I), $\Delta\omega/2\pi = 5$~MHz (II), and $\Delta\omega/2\pi = 10$~MHz (III). The first phonon mode is fixed at $\omega_1/2\pi = 10$~MHz (dashed red line). Green bars indicate the optimal configuration maximizing $E_N$. The corresponding value of $E_N$ is indicated below each configuration. (d) Bipartite entanglement $E_N^{p_+p_-}$ and (e) genuine tripartite entanglement $\mathcal{R}_\tau^{\min}(p_+, p_-, b_1)$ as continuous functions of the phonon frequency spacing $\Delta\omega$ for $k = 2, 6, 12, 25, 50$. The remaining parameters are given in the main text.}}
	\label{fig:2}
	\end{figure*}

	 The physical mechanism underlying the phenomenon of polariton-polariton entanglement is based on the phonon-mediated scattering of the driving field. The $k$-phonon modes generate Stokes ($\omega_0 - \omega_1$) and anti-Stokes ($\omega_0 + \omega_1$) sidebands from the intracavity photons at frequency $\omega_0$. By tuning the polariton detunings such that $\Delta_+ \simeq \omega_1$ and $\Delta_- \simeq -\omega_1$, the lower polariton $p_-$ couples to the Stokes sideband through a parametric down-conversion interaction that builds quantum correlations with the phonon ensemble, while the upper polariton $p_+$ couples to the anti-Stokes sideband through a beam-splitter interaction that maps these correlations onto $p_+$. The combined action of both processes results in stationary entanglement between the two polaritons. For $k=2$, for example, two parallel scattering pathways contribute to the correlation transfer, in a manner reminiscent of the dual phonon channels available in exciton-polariton microcavities where heavy-hole and light-hole polaritons couple to distinct GHz mechanical modes~\cite{C1}. In the parameter regime explored here, this multimode configuration provides a collective mediation mechanism in which each additional phonon mode opens a new scattering pathway. The quantitative impact on entanglement is analyzed in detail below. Additionally, in the regime where anti-Stokes scattering prevails, the phonon modes are driven toward their ground state, showing that the same configuration simultaneously supports mechanical cooling. Since each phonon mode contributes most efficiently when its frequency falls near the polariton resonance, the spectral distribution of the phonon modes plays a crucial role in determining the achievable entanglement. When several phonon modes with comparable coupling strengths participate, their scattering processes contribute in parallel to the polariton-polariton correlation transfer.	
	 	
Fig.~\ref{fig:2} shows the dependence of the two polariton entanglement on the phonon frequency spacing $\Delta\omega = \omega_{j+1} - \omega_j$ for three representative values $\Delta\omega/2\pi = 2$, $5$, and $10$~MHz, with the first phonon mode fixed at $\omega_1/2\pi = 10$~MHz. We consider three configurations with $k=2$ [Fig.~\ref{fig:2}(a)], $k=6$ [Fig.~\ref{fig:2}(b)], and $k=12$ [Fig.~\ref{fig:2}(c)] in order to identify the optimal frequency spacing as the number of phonon channels increases. The $j$-th phonon mode has frequency $\omega_j = \omega_1 + (j-1)\Delta\omega$. For instance, at $k=2$ the second mode corresponds to $\omega_2/2\pi = 12$, $15$, and $20$~MHz for the three spacings considered above. We note that Zuo \textit{et al.} \cite{A3} considered only two phonon modes with $\omega_2 = 3\omega_1$, corresponding to a fixed spacing of $\Delta\omega/2\pi = 20$~MHz. Here, we systematically explore smaller spacings and extend the analysis well beyond $k=2$, which allows us to determine the spacing that maximizes entanglement while keeping all phonon frequencies in the megahertz range, well below the magnon frequency in the gigahertz range, thereby ensuring that the magnomechanical interaction remains in the dispersive regime.
	Fig.~\ref{fig:2}(a)-(c) shows that among the three configurations considered, $\Delta\omega/2\pi = 2$~MHz consistently yields the highest entanglement for $k = 2, 6, 12$. This behaviour arises because a smaller spacing keeps all phonon frequencies closer to the polariton resonance at $\omega_1$, where the phonon-mediated scattering is most efficient. A larger spacing pushes the higher-frequency modes further from the polariton resonance, reducing their scattering efficiency. This effect becomes increasingly significant as $k$ grows: for $k=2$ the enhancement is moderate ($E_N = 0.135$ versus $0.110$), whereas for $k=12$ it is substantial ($E_N = 0.207$ versus $0.120$), since a large spacing places the highest mode at $\omega_{12}/2\pi = 120$~MHz, further from the polariton resonance, while a small spacing keeps it at $\omega_{12}/2\pi = 32$~MHz.
	
	To determine whether $\Delta\omega$ could be reduced even further, we scan $\Delta\omega$ continuously from zero up to $12$~MHz in Fig.~\ref{fig:2}(d) and (e), at the fixed hybridization $\theta = 0.4\pi$ used in Fig.~\ref{fig:2}(a)-(c). For small $k$, the bipartite entanglement is maximized near $\Delta\omega = 0$ and decreases smoothly with $\Delta\omega$. The genuine tripartite entanglement $\mathcal{R}_\tau^{\min}(p_+,p_-,b_1)$ shown in panel~(e) is identically zero at $\Delta\omega = 0$ for all $k$ tested, and becomes non-zero only above a $k$-dependent activation threshold: $\Delta\omega/2\pi \simeq 150$~kHz for $k=2$, $\simeq 500$~kHz for $k=6$, $\simeq 700$~kHz for $k=12$, $\simeq 900$~kHz for $k=25$, and $\simeq 1.2$~MHz for $k=50$. Beyond this threshold, $\mathcal{R}_\tau^{\min}$ rises, peaks at $\Delta\omega/2\pi \simeq 2$ to $4$~MHz, and then decreases. The choice $\Delta\omega/2\pi = 2$~MHz used in the rest of the manuscript is therefore a physical compromise. It is small enough to keep the bipartite entanglement appreciable and, being well above the activation thresholds for all $k$ considered, large enough to activate the genuine tripartite correlations. It also keeps all phonon frequencies deep in the dispersive regime ($\omega_j \ll \omega_c$) \cite{A7}.
		
	\begin{figure}[t]
		\centering
		\includegraphics[scale=0.23]{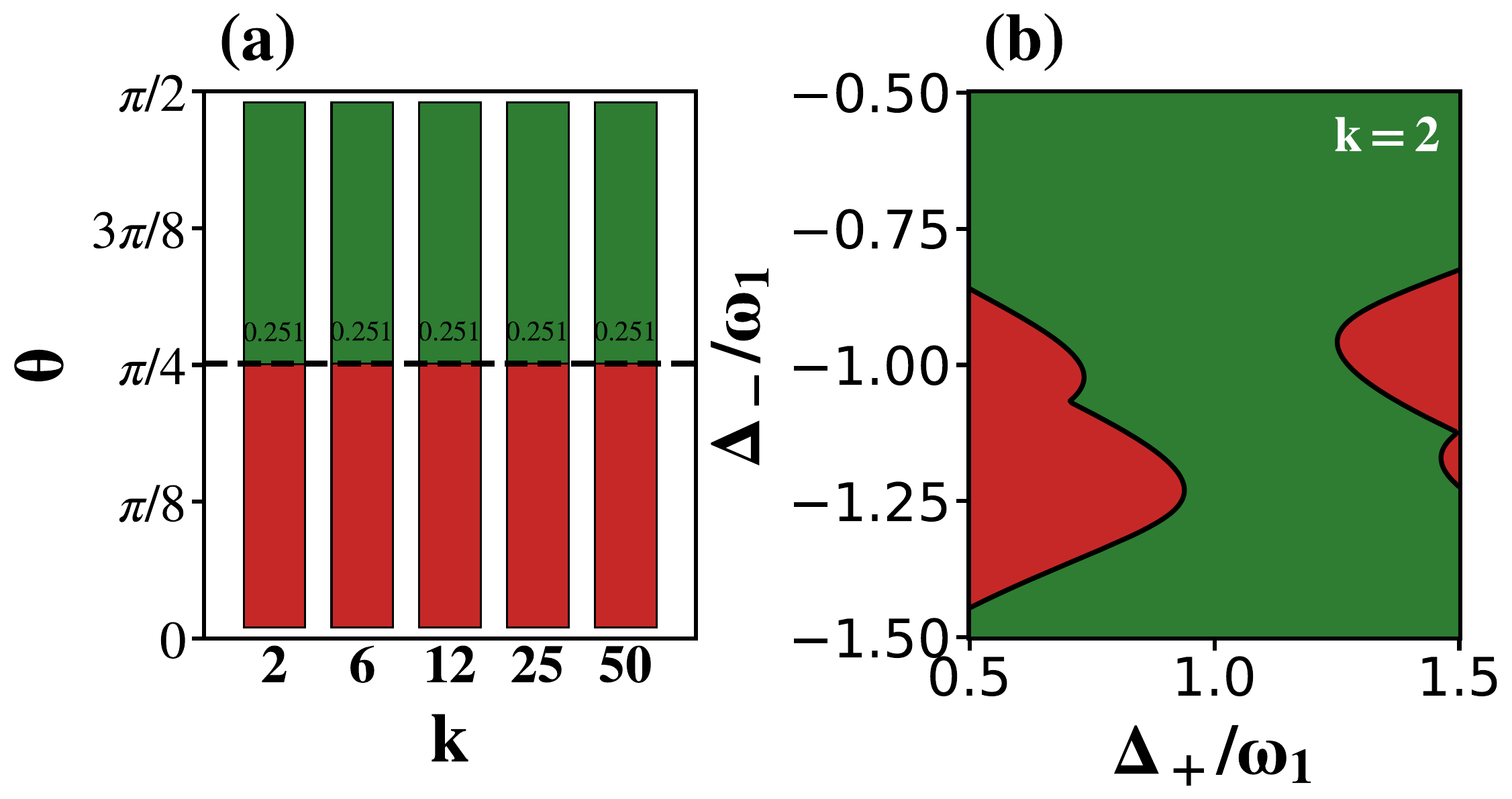}
		\captionsetup{justification=RaggedRight, singlelinecheck=false}
		\vspace{-1.2em} 
		\caption{{\small Stability diagram of the system. (a) Stable (green) and unstable (red) regions as a function of $\theta$ for $k = 2, 6, 12, 25$, and $50$ phonon modes at $\Delta_+ = -\Delta_- = \omega_1$. The dashed line marks the stability boundary $\theta_{\rm crit} \simeq \pi/4$, which is independent of $k$. (b) Stability map in the detuning plane $(\Delta_+/\omega_1,\,\Delta_-/\omega_1)$ for $k=2$ at $\theta = 0.4\pi$. The parameters are listed in Tab.~\eqref{tab:1}.}}
		\label{fig:02}
	\end{figure}
	
	We employ the experimentally feasible parameters listed in Table~\ref{tab:1}. The optimal resonance condition is set as $\Delta_+ = -\Delta_- = \omega_1$. For the mixing angle $\theta = 0.4\pi$, the magnon frequency and the cavity--magnon coupling are determined by simultaneously solving $\theta = \frac{1}{2}\arctan(2g/\Delta_{ac})$ and $\sqrt{(\omega_a - \omega_c)^2 + 4g^2} = 2\omega_1$, yielding $g/2\pi \simeq 5.88$~MHz and $\omega_c/2\pi \simeq 10.016$~GHz. Experimentally, the two polariton frequencies can be continuously tuned either via the bias magnetic field $H_0$, which adjusts the detuning $\Delta_{ac}$, or via the position of the YIG sphere inside the cavity, which controls~$g$. For a YIG sphere of $250~\mu$m diameter, the bare magnomechanical coupling is inherently weak~\cite{A2}, and an intense microwave drive is required to significantly enhance the effective coupling to $\mathcal{G}_j/2\pi = 2$~MHz, corresponding to a bias field $B_0 = 2.12 \times 10^{-5}$~T, with $G_{0j}/2\pi = 0.2$~Hz under the conditions $\Delta_+ = -\Delta_- = \omega_1$ and $\theta = 0.4\pi$, which yields a drive power $\mathcal{P} = 2.64$~mW \cite{B1}. The stability of the system is guaranteed by the negative real parts of all eigenvalues of the drift matrix $\boldsymbol{\Lambda}$ for every parameter set considered, as shown in Fig.~\ref{fig:02}.
	
	\begin{table}[h]
	\caption{Experimentally feasible parameters~\cite{A1,A2}.}
	\label{tab:1}
	\begin{ruledtabular}
		\begin{tabular}{lcc}
			\textrm{Parameter} & \textrm{Symbol} & \textrm{Value} \\
			\colrule
			Cavity frequency & $\omega_a/2\pi$ & $10$~GHz \\
			Magnon frequency & $\omega_c/2\pi$ & $10.016$~GHz \\
			Phonon frequency & $\omega_1/2\pi$ & $10$~MHz \\
			Phonon frequency spacing & $\Delta\omega/2\pi$ & $2$~MHz \\
			Cavity decay rate & $\kappa_a/2\pi$ & $1$~MHz \\
			Magnon decay rate & $\kappa_c/2\pi$ & $1$~MHz \\
			Mechanical damping rate & $\kappa_j/2\pi$ & $10^2$~Hz \\
			Cavity--magnon coupling & $g/2\pi$ & $5.88$~MHz \\
			Bath temperature & $T$ & $10$~mK \\
		\end{tabular}
	\end{ruledtabular}
\end{table}
	
	\begin{figure}[h]
		\centering
		\includegraphics[scale=0.335 ]{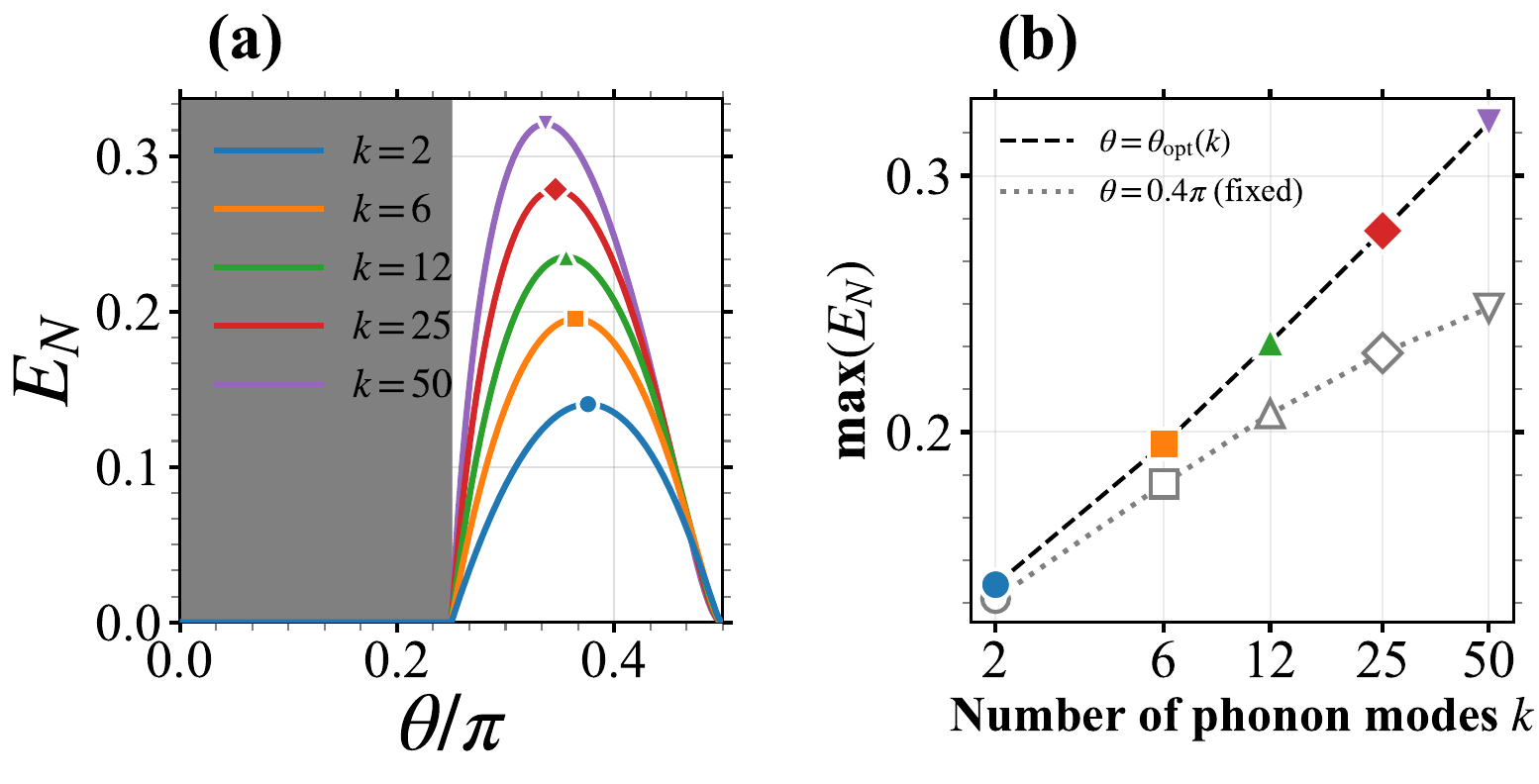}
		\captionsetup{justification=RaggedRight, singlelinecheck=false}
		\vspace{-1.2em} 
		\caption{{\small (a) Stationary polariton-polariton entanglement $E_N$ as a function of the mixing angle $\theta$ for $k = 2$, $6$, $12$, $25$, and $50$ phonon modes. The gray region denotes the dynamically unstable regime. Markers indicate the maximum $E_N$ for each $k$. (b) Maximum entanglement $E_N^{\mathrm{max}}$ as a function of the number of phonon modes $k$. The parameters are the same as in Fig.~\ref{fig:2}.}}
		\label{fig:3}
	\end{figure}
	
	Fig.~\ref{fig:3}(a) shows the stationary polariton entanglement $E_N$ as a function of the mixing angle $\theta$ for $k = 2$, $6$, $12$, $25$, and $50$ phonon modes. The system is dynamically unstable for $\theta/\pi \lesssim 0.25$, a regime where the Stokes and anti-Stokes scattering rates are nearly balanced and no stationary entanglement can be sustained. Beyond this threshold, $E_N$ exhibits a pronounced peak whose amplitude grows monotonically with $k$: $E_N^{\mathrm{max}} \simeq 0.14$ ($k=2$), $0.20$ ($k=6$), $0.24$ ($k=12$), $0.28$ ($k=25$), and $0.33$ ($k=50$), representing more than a twofold enhancement relative to the $k=2$ case. 
	Fig.~\ref{fig:3}(b) shows $E_N^{\mathrm{max}}$ as a function of $k$ for the optimized angle $\theta_{\mathrm{opt}}(k)$ (dashed) and for the fixed hybridization $\theta = 0.4\pi$ used in Fig.~\ref{fig:2} (dotted). Both curves grow monotonically with $k$, confirming the robustness of the collective phonon-mediated enhancement. At $k=50$, optimizing $\theta$ yields $E_N^{\mathrm{max}} \simeq 0.32$, about $1.3$ times larger than the value $\simeq 0.25$ obtained with $\theta = 0.4\pi$.
	This monotonic growth confirms that the phonon modes serve as parallel mediation channels through which quantum correlations are collectively distributed between the two polaritons. Notably, the optimal mixing angle shifts from $\theta_{\mathrm{opt}}/\pi \simeq 0.376$ at $k=2$ toward $\theta_{\mathrm{opt}}/\pi \simeq 0.337$ at $k=50$, approaching the instability boundary at $\theta = \pi/4$. This trend reflects the collective strengthening of the anti-Stokes process by the multi-phonon ensemble, which relaxes the requirement for a large asymmetry between the two scattering channels to maintain stability. 
	Experimentally, the resonance condition $\Delta_+ = -\Delta_- = \omega_1$ is maintained by adjusting the cavity--magnon coupling $g$ via the YIG sphere position and the magnon frequency $\omega_c$ via the bias field $H_0$. For the fixed synchronization $\theta = 0.4\pi$ used in Fig.~\ref{fig:2}, this corresponds to the values given in Table~\ref{tab:1}. Since $\theta_{\mathrm{opt}}$ varies with $k$, the required coupling $g/2\pi$ and magnon frequency $\omega_c/2\pi$ change accordingly. Going from $k=2$ to $k=50$, $g/2\pi$ increases from $7.03$ to $8.54$~MHz, while $\omega_c/2\pi$ decreases from $10.014$ to $10.010$~GHz.
	These small adjustments are well within the experimentally accessible range, where cavity--magnon couplings up to $g/2\pi = 143$~MHz have been demonstrated for different YIG sphere sizes~\cite{A11}. 
	This scaling behavior is a key signature distinguishing the present multimode scheme from conventional single-mode protocols.
	  
\begin{figure}[h]
\centering
\includegraphics[scale=0.27]{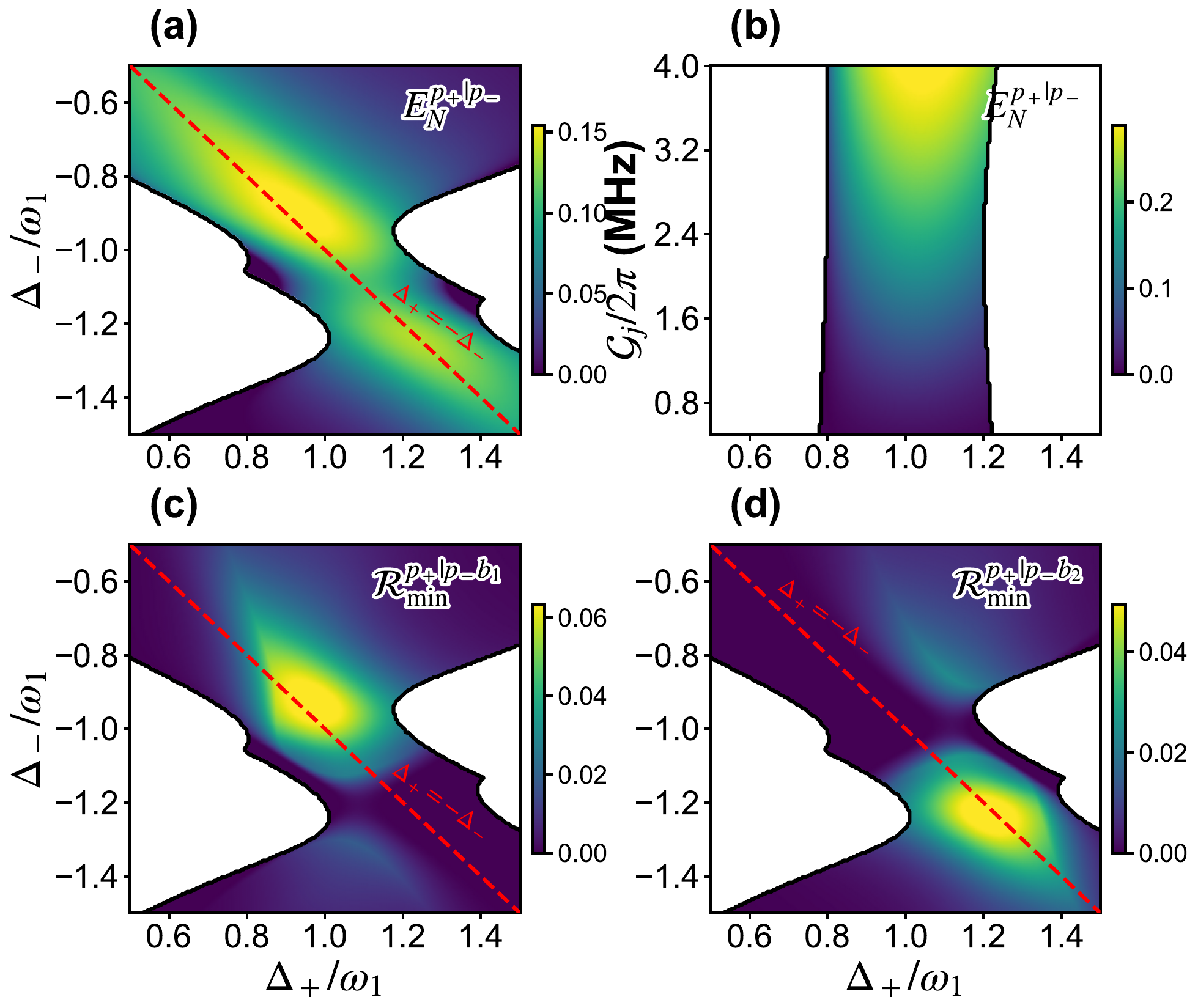}
\captionsetup{justification=RaggedRight, singlelinecheck=false}
\vspace{-1.2em} 
\caption{{\small (a) Stationary polariton-polariton entanglement $E_N$ as a function of the detunings $\Delta_+/\omega_1$ and $\Delta_-/\omega_1$ for $k=2$ phonon modes. (b) $E_N$ as a function of $\Delta_+/\omega_1$ and the effective magnomechanical coupling $\mathcal{G}_j/2\pi$ with $\Delta_- = -\omega_1$. (c,d) Minimum residual contangle $\mathcal{R}_{\min}$ quantifying genuine tripartite entanglement among the two polaritons and the first (c) or second (d) phonon mode. White regions denote parameter sets where the system is dynamically unstable. The parameters are the same as in Fig.~\ref{fig:2}.}}		
		\label{fig:4}
	\end{figure}

	Fig.~\ref{fig:4} presents the entanglement landscape for $k=2$ phonon modes at the optimal mixing angle $\theta_{\mathrm{opt}}/\pi \simeq 0.376$ identified in Fig.~\ref{fig:3}. Panel~(a) shows the polariton-polariton entanglement $E_N^{p_+p_-}$ as a function of the detunings $\Delta_+$ and $\Delta_-$. The entanglement is maximized along the values $\Delta_+ = -\Delta_-$ (red dashed line), peaking near $|\Delta_\pm|/\omega_1 \simeq 1$, confirming the optimal resonance condition $\Delta_+ = -\Delta_- \simeq \omega_1$ where both Stokes and anti-Stokes scattering processes are activated simultaneously. Deviations from this condition rapidly suppress the entanglement, while high values of $|\Delta_\pm|$ drive the system into dynamically unstable regions (white areas). Panel~(b) displays $E_N^{p_+p_-}$ as a function of the effective magnomechanical coupling $\mathcal{G}_j$ and detuning $\Delta_+$ at the optimal condition $\Delta_- = -\omega_1$. Throughout this analysis, we assume comparable bare magnetostrictive couplings $G_{0j} \simeq G_0$ for the selected phonon modes. Since $\mathcal{G}_j \propto G_{0j}\langle c\rangle$, the microwave drive uniformly enhances all couplings to a common value $\mathcal{G}$, allowing us to isolate the role of the number of phonon modes $k$.
The entanglement increases with $\mathcal{G}_j$ and reaches its maximum well within the stable regime. The underlying mechanism, in which Stokes scattering generates quantum correlations between the lower polariton $p_-$ and the phonon modes, while anti-Stokes scattering transfers these correlations to the upper polariton $p_+$, is consistent with the single-phonon protocol of Ref.~\cite{A3}. As in that case, the achievable entanglement is not constrained by the stability boundary, in contrast to the schemes reported in Refs.~\cite{B1,B3}.

In the present multimode scenario, however, the interaction is mediated by multiple phonon modes that act as parallel scattering channels, leading to an accumulation of correlation transfer processes. This results in a more efficient transfer of quantum correlations between the two polaritons, enabling higher levels of entanglement compared to the single-phonon case. Panels~(c) and~(d) of Fig.~\ref{fig:4} show the minimum residual contangle $\mathcal{R}_{\tau}^{\min}$, which quantifies genuine tripartite entanglement among the two polaritons and the first or second phonon mode, respectively. The tripartite entanglement involving $b_1$ peaks near $|\Delta_\pm|/\omega_1 \simeq 1.0$, while that involving $b_2$ peaks near $|\Delta_\pm|/\omega_1 \simeq 1.2$, consistent with the respective phonon frequencies $\omega_1/2\pi = 10$~MHz and $\omega_2/2\pi = 12$~MHz. This frequency-selective response indicates that each phonon mode mediates tripartite correlations most efficiently when the polariton splitting matches its mechanical resonance.
	Such mode-resolved mediation is absent in single-phonon schemes and represents a distinctive feature of the multimode extension, highlighting its potential for frequency-selective and scalable quantum correlation engineering.
		
		\begin{figure}[h]
		\centering
		\includegraphics[scale=0.3]{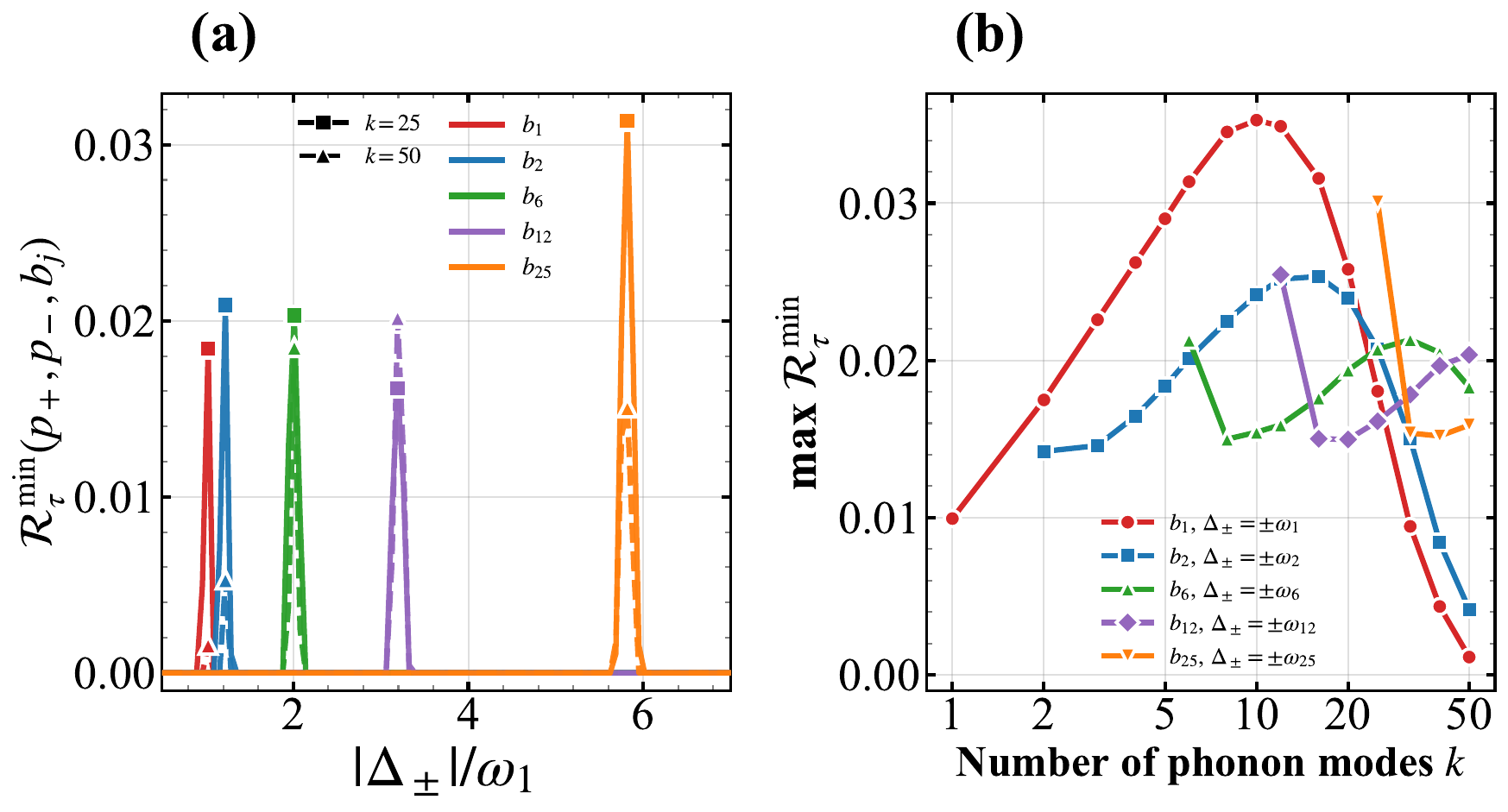}
		\captionsetup{justification=RaggedRight, singlelinecheck=false}
		\vspace{-1.2em} 
		\caption{{\small Genuine tripartite entanglement $\mathcal{R}_\tau^{\min}(p_+,p_-,b_j)$ between the two polaritons and the $j$-th phonon mode, for $j = 1, 2, 6, 12, 25$, with $\omega_j = \omega_1 + (j-1)\Delta\omega$ and $\Delta\omega/2\pi = 2$~MHz. (a) As a function of $|\Delta_\pm|/\omega_1$ at $k = 25$ (solid lines) and $k = 50$ (dashed lines). The markers indicate the position of the maximum for each $k$. (b) As a function of $k$ at the matched resonance $\Delta_\pm = \pm \omega_j$. In both panels, $\theta$ is re-optimized to maximize the bipartite polariton entanglement at the corresponding detuning. The parameters are the same as in Fig.~\ref{fig:2}.}}
		\label{fig:04}
	\end{figure}
	
To confirm this frequency-selective behavior across the whole mode ensemble and to explore its scaling with $k$, we present in Fig.~\ref{fig:04} the tripartite entanglement $\mathcal{R}_\tau^{\min}(p_+,p_-,b_j)$ for $j = 1, 2, 6, 12, 25$, with $\theta$ re-optimized at each detuning to maximize the bipartite polariton entanglement. Panel~(a) shows $\mathcal{R}_\tau^{\min}$ as a function of $|\Delta_\pm|/\omega_1$ at $k = 25$ (solid) and $k = 50$ (dashed). Each mode exhibits a narrow peak centered at $|\Delta_\pm| = \omega_j$ and is essentially zero elsewhere, extending the mode-selective response observed in Fig.~\ref{fig:4}(c),(d) to higher-order phonons. Panel~(b) shows $\mathcal{R}_\tau^{\min}$ as a function of $k$ at the matched resonance $|\Delta_\pm| = \omega_j$. Each curve is non-monotonic and reaches an optimum at a finite value of $k$, revealing a trade-off between the collective enhancement provided by additional modes and the dilution of correlations across the mode ensemble. This behavior has no counterpart in the single-phonon protocol and provides a quantitative signature of the multimode scheme.

	\begin{figure}[h]
		\centering
		\includegraphics[scale=0.265]{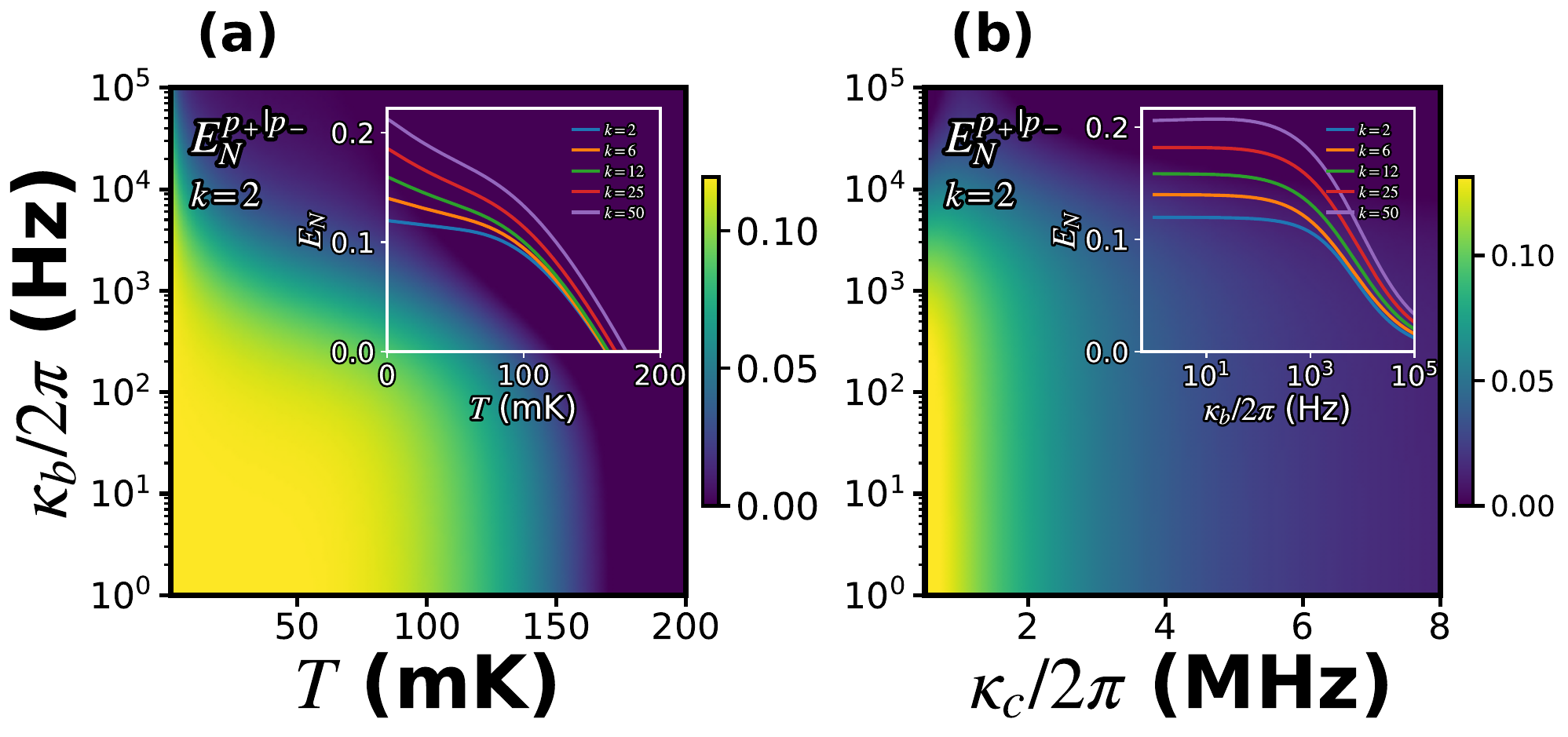}
		\captionsetup{justification=RaggedRight, singlelinecheck=false}
		\vspace{-1.2em} 
		\caption{{\small Robustness of the polariton entanglement for $k=2$ phonon modes. (a) $E_N^{p_+p_-}$ versus bath temperature $T$ and mechanical damping rate $\kappa_b/2\pi$. (b) $E_N^{p_+p_-}$ versus magnon dissipation rate $\kappa_c/2\pi$ and $\kappa_b/2\pi$. White regions denote dynamically unstable parameter sets. Insets: $E_N$ for $k=2$, $6$, $12$, $25$, and $50$ phonon modes versus $T$ at fixed $\kappa_b/2\pi = 100$~Hz (a), and versus $\kappa_b/2\pi$ at fixed $\kappa_c/2\pi = 1$~MHz~(b). The other parameters are the same as in Fig.~\ref{fig:2}.}}
			\label{fig:5}
	\end{figure}
	
	Fig.~\ref{fig:5} shows the robustness of the polariton-polariton entanglement against thermal bath and dissipation for $k=2$ phonon modes. Panel~(a) shows $E_N^{p_+p_-}$ as a function of the bath temperature $T$ and the mechanical damping rate $\kappa_b$. The entanglement persists up to $T \simeq 150$~mK when mechanical damping is low, and survives for $\kappa_b$ up to $10^4$~Hz at millikelvin temperatures, confirming experimental feasibility within current cryogenic setups. Panel~(b) shows $E_N^{p_+p_-}$ as a function of the dissipation rates $\kappa_c$ and $\kappa_b$, with the cavity dissipation rate set at $\kappa_a/2\pi = 1$~MHz. The entanglement reaches its maximum when $\kappa_c \neq \kappa_a$, demonstrating that an asymmetry between the two dissipation channels is advantageous. When $\kappa_a = \kappa_c$, the two polaritons experience identical effective dissipation, leading to a destructive interference between the dissipation pathways that suppresses the asymmetry between the Stokes and anti-Stokes processes required for entanglement generation. Breaking this symmetry by tuning $\kappa_c$ away from $\kappa_a$ restores the imbalance and enhances the entanglement. The insets compare the thermal and dissipative robustness across different $k$, showing that configurations with larger $k$ consistently yield stronger entanglement and improved thermal robustness.

	   \begin{figure}[t]
		\centering
		\includegraphics[scale=0.16]{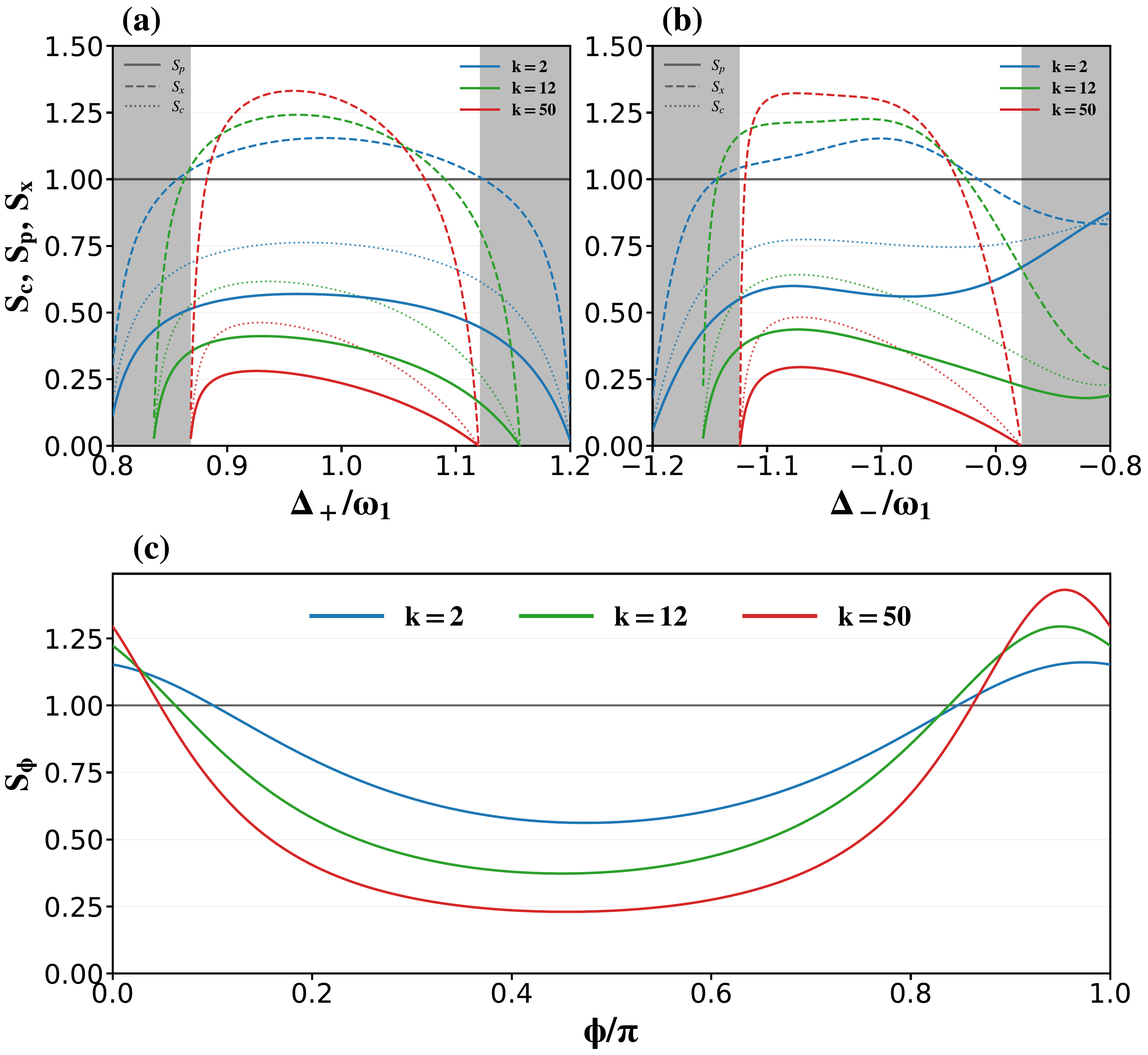}
		\captionsetup{justification=RaggedRight, singlelinecheck=false}
		\vspace{-1.2em} 
		\caption{{\small Quantum synchronization between the two polariton modes for $k=2$, $12$, and $50$ phonon modes. (a) Complete synchronization $S_c$ (dotted), amplitude synchronization $S_x$ (dashed), and phase synchronization $S_p$ (solid) as a function of $\Delta_+/\omega_1$ at $\Delta_- = -\omega_1$. (b) Same quantities as a function of $\Delta_-/\omega_1$ at $\Delta_+ = \omega_1$. (c) Generalized synchronization $S_\phi$ as a function of the quadrature angle $\phi/\pi$ at the resonance condition $\Delta_\pm = \pm\omega_1$. Gray shaded regions denote unstable parameter sets. The dotted horizontal line marks the vacuum level $S = 1$. All parameters are the same as in Fig.~\ref{fig:2}, with $\theta = \theta_{\mathrm{opt}}(k)$ for each curve.}}
		\label{fig:6}
	\end{figure}
	
	Finally, having established the entanglement properties of the system, we now investigate the quantum correlations between the two polaritons through their synchronization behavior. Fig.~\ref{fig:6} shows the quantum synchronization measures $S_c$, $S_p$, and $S_x$ between the two polariton modes as a function of the detunings $\Delta_+$ [panel~(a)] and $\Delta_-$ [panel~(b)] for $k = 2$, $12$, and $50$, together with the generalized synchronization measure $S_\phi$ [panel~(c)]. All measures reach their maximum near the optimal resonance condition $|\Delta_\pm|/\omega_1 \simeq 1$, confirming that the same Stokes/anti-Stokes scattering mechanism responsible for entanglement also drives synchronization. We observe that $S_c$ is consistently larger than $S_p$ in all configurations, indicating that the amplitude quadrature difference $\langle\tilde{X}_-^2\rangle$ is smaller than its phase counterpart $\langle\tilde{P}_-^2\rangle$, so that the two polaritons track each other more closely in amplitude than in phase. Increasing $k$ reduces both $S_c$ and $S_p$, with $S_c$ decreasing from $\simeq 0.76$ for $k=2$ to $\simeq 0.40$ for $k=50$. This is in stark contrast to the entanglement $E_N$, which increases monotonically with $k$, as shown in Fig.~\ref{fig:3}(b). 
	
Interestingly, the amplitude synchronization $S_x$ exceeds unity for all $k$ tested, ranging from $\simeq 1.15$ ($k=2$) to $\simeq 1.30$ ($k=50$). This indicates that $\langle\tilde{X}_-^2\rangle$ falls below the vacuum level $1/2$, revealing collective squeezing in the amplitude quadrature of the polariton difference mode. Panel~(c) confirms this quadrature selectivity by displaying the generalized synchronization $S_\phi$, which recovers $S_x$ at $\phi = 0$ and $S_p$ at $\phi = \pi/2$. Its maximum reaches $\simeq 1.16$, $1.29$, and $1.43$ for $k = 2$, $12$, and $50$, respectively. Notably, both $S_x$ and $\max S_\phi$ grow with $k$, showing that multimode mediation enhances collective amplitude squeezing without an external squeezing drive. This can be understood from the fact that additional phonon channels enhance quantum correlations between the two polaritons while simultaneously introducing collective noise that degrades the precision of their mutual tracking in the phase quadrature. Specifically, each phonon channel contributes correlated fluctuations that strengthen entanglement, but their contributions add incoherently in the phase direction, causing $\langle \tilde{P}_-^2 \rangle$ to increase with $k$, while $\langle \tilde{X}_-^2 \rangle$ benefits from the constructive accumulation of amplitude correlations and decreases. Consequently, the system exhibits a fundamental trade-off between these two quantum resources, where enhancing entanglement comes at the cost of reduced phase synchronization while amplitude squeezing is simultaneously enhanced. This interplay is absent in single-mode systems and emerges as a distinctive feature of multimode-mediated hybrid quantum platforms.

Our protocol is particularly relevant for platforms that naturally host multiple mechanical modes, such as optomechanical crystal arrays, exciton-optomechanical microcavities, and plasmon-optomechanical photonic crystals. Furthermore, we comment on the robustness of the proposed mechanism against experimental imperfections. The entanglement generation relies on the resonance condition $\Delta_{+} \simeq \omega_1$ and $\Delta_{-} \simeq -\omega_1$, which can be affected by detuning fluctuations. However, due to the presence of multiple phonon modes, the system exhibits an inherent tolerance to such imperfections, as nearby modes can still contribute effectively to the scattering processes. Thermal effects, characterized by the mean phonon occupation numbers $\bar{n}_{b_j}$, tend to degrade quantum correlations. Nevertheless, the collective enhancement from multiple modes partially compensates for this degradation, allowing entanglement to persist at finite temperatures. The proposed scheme also tolerates deviations of individual phonon frequencies $\omega_j$ from their ideal values, since the polariton entanglement varies smoothly with such deviations. These considerations indicate that the proposed scheme is robust and compatible with current experimental capabilities.

	\section{Conclusion}
	\label{sec:conclusion}

    In summary, we have presented a mechanism for generating and controlling quantum correlations in cavity magnomechanical systems through multimode phonon mediation. By coupling cavity--magnon polaritons to multiple vibrational modes, we have shown that the phonon modes act as parallel scattering channels that collectively enhance the transfer of quantum correlations between the polaritons. In contrast to conventional single-mode schemes, this multimode configuration gives rise to a monotonic increase of steady-state entanglement with the number of phonon modes in the parameter regime explored here.
Furthermore, we have demonstrated that quantum synchronization between the polariton modes originates from the same underlying Stokes and anti-Stokes scattering processes responsible for entanglement generation, yet exhibits an opposite dependence on the number of phonon modes in the phase quadrature, while the amplitude synchronization $S_x$ exceeds unity and grows with $k$, revealing collective amplitude squeezing without an external squeezing drive. This reveals a nontrivial interplay between synchronization and entanglement, highlighting a rich quadrature structure of multimode hybrid systems. We also find that the tripartite entanglement among the two polaritons and each phonon mode peaks at a frequency determined by the individual mechanical resonance, revealing a mode-resolved mediation.
Overall, these results indicate that multimode phonon mediation provides a viable and experimentally relevant route toward collective quantum resource engineering in hybrid systems, with potential applications in quantum information processing, quantum networks, and scalable hybrid quantum technologies. Future investigations may explore the impact of nonlinearities, disorder in phonon spectra, and realistic implementations in multimode cavity architectures, paving the way toward practical realizations of multimode-controlled quantum dynamics.

	\section*{Acknowledgments}
	
	Z. Imara acknowledges the UM6P Vanguard Center for hospitality and support during the completion of this work. 
	Jia-Xin Peng is supported by National Natural Science Foundation of China (Grant No.~12504566), the Basic Research Program of Jiangsu (Grant No.~BK20250947), Natural Science Foundation of the Jiangsu Higher Education Institutions (Grant No.~25KJB140013), and  Natural Science Foundation of Nantong City (Grant No.~JC2024045). S. K. Singh gratefully acknowledges the high-performance computing facilities provided by Akal University, Punjab, India.

\end{document}